\documentclass{article}

\usepackage{PRIMEarxiv}

\usepackage[utf8]{inputenc} % allow utf-8 input
\usepackage[T1]{fontenc}    % use 8-bit T1 fonts
\usepackage[hypertexnames=false]{hyperref}       % hyperlinks
\usepackage{url}            % simple URL typesetting
\usepackage{booktabs}       % professional-quality tables
\usepackage{amsfonts}       % blackboard math symbols
\usepackage{nicefrac}       % compact symbols for 1/2, etc.
\usepackage{microtype}      % microtypography
\usepackage{lipsum}
\usepackage{fancyhdr}       % header
\usepackage{graphicx}       % graphics
\usepackage{multirow}
\usepackage{array}
\usepackage{inconsolata}
\usepackage{placeins}
\usepackage{algorithm}
\usepackage{algpseudocode}
\usepackage{amsmath}
\usepackage{amssymb,enumitem}
\graphicspath{{media/}}     

\title{Anonymization-Enhanced Privacy Protection for Mobile GUI Agents: Available but Invisible

}

\author{
  Lepeng Zhao  \\
  Tsinghua University \\
  \texttt{zhaolp22@mails.tsinghua.edu.cn} \\
   \And
  Zhenhua Zou \\
  Tsinghua University \\
  \texttt{zou-zh21@mails.tsinghua.edu.cn} \\
   \And
  Shuo Li \\
  Tsinghua University \\
  \texttt{shuo-li22@mails.tsinghua.edu.cn} \\
   \And
  Zhuotao Liu\thanks{Corresponding author.} \\
  Tsinghua University \\
  \texttt{zhuotaoliu@tsinghua.edu.cn} \\
}

\begin{document}
\maketitle

\begin{abstract}
Mobile Graphical User Interface (GUI) agents have demonstrated strong capabilities in automating complex smartphone tasks by leveraging multimodal large language models (MLLMs) and system-level control interfaces. However, this paradigm introduces significant privacy risks, as agents typically capture and process entire screen contents, thereby exposing sensitive personal data such as phone numbers, addresses, messages, and financial information. Existing defenses either reduce UI exposure, obfuscate only task-irrelevant content, or rely on user authorization, but none can protect task-critical sensitive information while preserving seamless agent usability.

We propose an anonymization-based privacy protection framework that enforces the principle of \emph{available-but-invisible} access to sensitive data: sensitive information remains usable for task execution but is never directly visible to the cloud-based agent. Our system detects sensitive UI content using a PII-aware recognition model and replaces it with deterministic, type-preserving placeholders (e.g., \texttt{PHONE\_NUMBER\#a1b2c}) that retain semantic categories while removing identifying details. A layered architecture—comprising a PII Detector, UI Transformer, Secure Interaction Proxy, and Privacy Gatekeeper—ensures consistent anonymization across user instructions, XML hierarchies, and screenshots, mediates all agent actions over anonymized interfaces, and supports narrowly scoped local computations when reasoning over raw values is necessary.

Extensive experiments on the AndroidLab and PrivScreen benchmarks show that our framework substantially reduces privacy leakage across multiple models while incurring only modest utility degradation, yielding a favorable privacy--utility trade-off compared with existing methods. Code available at: \url{https://github.com/one-step-beh1nd/gui_privacy_protection}

\vspace{0.5em}
\end{abstract}

\section{Introduction}

Mobile Graphical User Interface (GUI) agents~\cite{appagent, mobileagent, uitars2, autoglm, appcopilot} have recently demonstrated remarkable capabilities in automating complex tasks on smartphones, including app navigation, form filling, information retrieval, and multi-step interaction workflows. These capabilities are enabled by two key factors: (1) the rapid development of multimodal large language models (MLLMs)~\cite{qwen25vl}, which provide strong perceptual and reasoning abilities over screenshots, UI hierarchies, and visual layouts; and (2) long-standing system-level control interfaces such as Android Debug Bridge (ADB) and Accessibility Services, which make it possible for agents to programmatically observe and operate the device. Together, these two factors allow users to delegate tasks to GUI agents in a manner analogous to handing over control of their personal smartphone to another human operator.

\begin{figure}[htbp]
    \centering
    \includegraphics[width=\textwidth]{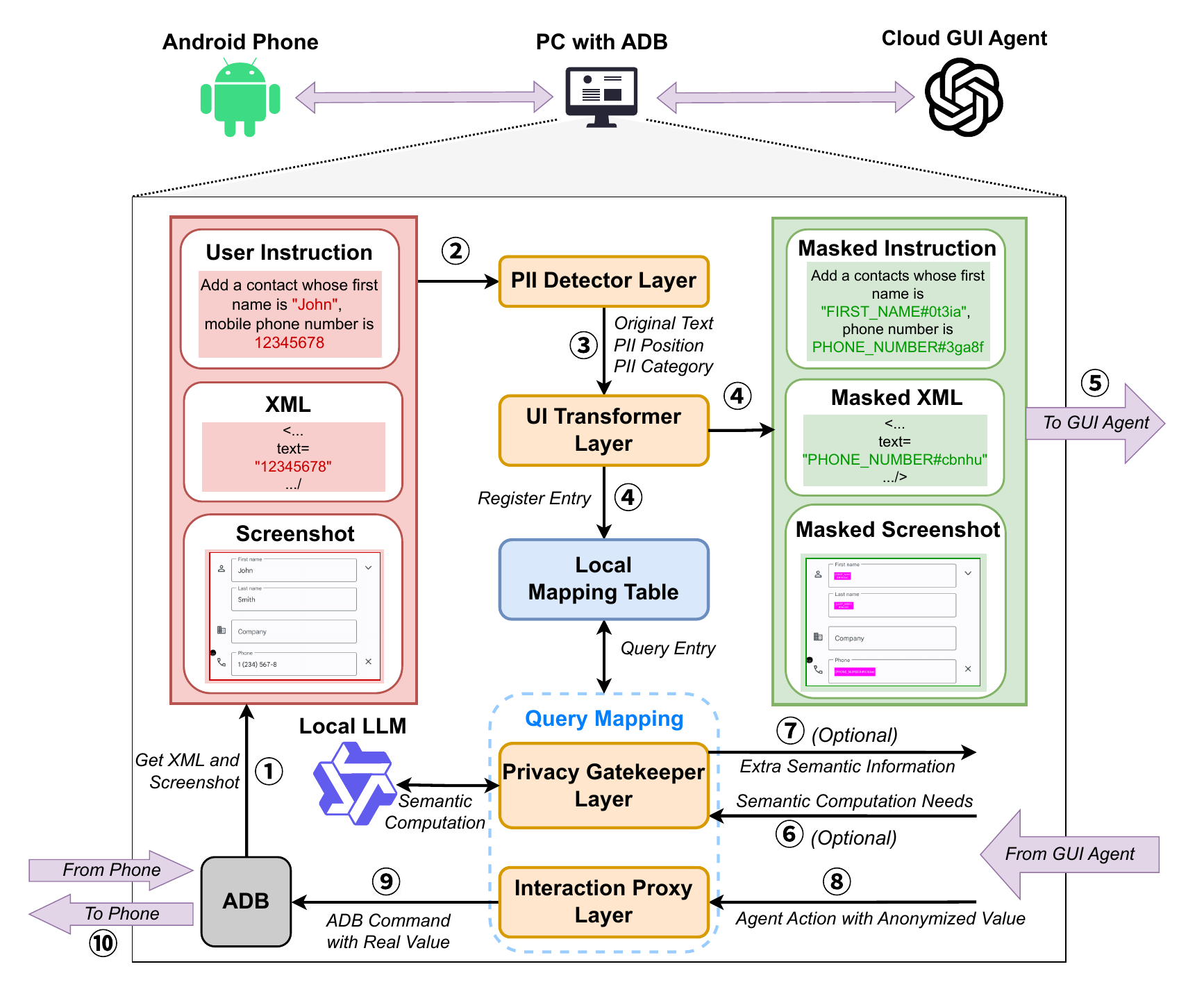}
    \caption{Overview of the proposed privacy protection framework for mobile GUI agents. The system inserts a trusted local privacy layer between the mobile phone and a cloud-based GUI agent. User instructions and UI states (XML hierarchies and screenshots) are first processed locally to detect sensitive content and replace it with type-preserving anonymized placeholders, producing an anonymized Virtual UI for agent reasoning. The cloud agent operates exclusively on this anonymized interface and issues actions based on placeholders. All actions are intercepted by a local interaction proxy, which resolves anonymized references and executes them on the phone using original values when necessary. For tasks requiring operations over raw sensitive data, a local privacy gatekeeper performs limited computation and returns only non-sensitive results to the agent.}
    \label{fig:overview}
\end{figure}

However, this paradigm also introduces significant privacy risks. To perceive and reason about the interface, existing GUI agents typically capture and process the entire screen content—either via screenshots or structured UI representations (e.g., XML view hierarchies). This leads to systematic over-collection of information, where sensitive user data such as phone numbers, addresses, chat messages, verification codes, and financial information are unnecessarily exposed to the agent or transmitted to remote servers. In practice, such data may also be persistently stored; for example, systems like AppCopilot~\cite{appcopilot} automatically cache extracted user information into internal user modules to support future tasks, further increasing the attack surface and long-term privacy risks.

Several prior works have attempted to mitigate privacy leakage in GUI agents and related systems, yet each follows a design paradigm that ultimately fails to protect sensitive information that is necessary for task execution. Existing approaches can be broadly categorized into four classes.
(1) \emph{Exposure reduction} methods, such as CORE~\cite{core}, reduce the amount of UI data sent to cloud-based models but still transmit raw sensitive content whenever it is task-relevant.
(2) \emph{Obfuscation-based} methods, such as DualTAP~\cite{dualtap}, perturb sensitive regions in the UI, but deliberately weaken perturbations for task-relevant private content to preserve utility, leaving such information partially recognizable.
(3) \emph{Access-control and authorization-based} methods, such as PrivWeb~\cite{privweb}, rely on user consent to regulate disclosure, but once authorization is granted, real sensitive values are directly exposed to the agent or remote services.
(4) \emph{Semantic replacement} methods, such as GUIGuard~\cite{guiguard}, attempt to mask PII with generic categories, yet the agent often loses the ability to use the masked PII (e.g., entering a phone number) because the connection between the masked UI and the original data is not functionally preserved.
\begin{table}[htbp]
\centering
\caption{Comparison of privacy protection and task usability across different GUI agent frameworks. The table specifies the \textbf{leakage conditions} for PII and whether the agent can \textbf{effectively utilize} task-relevant sensitive data while it remains protected. The symbol $\times$ indicates that the PII is effectively protected from cloud exposure.}
\label{tab:privacy-comparison}
\resizebox{\textwidth}{!}{
\begin{tabular}{@{}llll@{}}
\toprule
\textbf{Method} & \textbf{Task-Relevant PII} & \textbf{Task-Irrelevant PII} & \textbf{Protected Data Usability} \\ \midrule
CORE~\cite{core} & Exists in task-related blocks & Exists in task-related blocks & -- (Unprotected when using) \\
PrivWeb~\cite{privweb} & Following user authorization & Following user authorization & -- (Unprotected when using) \\
DualTAP~\cite{dualtap} & Insufficient image perturbation & $\times$ & Limited (Utility loss) \\
GUIGuard~\cite{guiguard} & $\times$ & $\times$ & Partial (Reasoning-based) \\ \midrule
\textbf{Ours} & \textbf{$\times$} & \textbf{$\times$} & \textbf{Yes (Proxy-mediated)} \\ \bottomrule
\end{tabular}
}
\end{table}
As summarized in Table~\ref{tab:privacy-comparison}, these approaches struggle to simultaneously achieve strong privacy guarantees and seamless usability. Crucially, they either fail to protect task-critical sensitive information or significantly disrupt the agent’s ability to \emph{utilize} that data for task completion. Even when data is "protected" (e.g., via blurring or generic masking), the resulting loss of information often renders the agent unable to ground its actions or handle sensitive values accurately.

In this work, we adopt a different perspective and draw inspiration from privacy-preserving computation, particularly the principle of making data \emph{available but invisible}. Our goal is to enable GUI agents to \emph{use} sensitive information to perform tasks without \emph{seeing} or accessing its real content.

This design introduces new technical challenges beyond simple anonymization. In particular, inconsistencies in PII detection across modalities (e.g., text vs.\ screenshots) and inconsistent placeholder assignment across time can break the agent’s ability to ground its actions in the interface, leading to execution failures. Our framework is therefore explicitly designed to ensure cross-modality and temporal consistency, so that the same semantic entity is always represented by the same anonymized identifier throughout a task session.

As shown in Figure~\ref{fig:overview}, our framework consists of four coordinated components: (1) a \emph{UI Sensitivity Analyzer} that detects sensitive elements using PII-aware models; (2) a \emph{UI Transformer} that performs deterministic, type-preserving anonymization while preserving layout and interaction affordances; (3) a \emph{Secure Interaction Proxy} that mediates and resolves all agent actions over anonymized content; and (4) a local \emph{Privacy Gatekeeper} that performs tightly scoped semantic operations over raw values when necessary and returns only minimal, non-revealing results to the agent. 

A session-scoped local mapping table is maintained within the trusted environment to record the correspondence between real entity values and their anonymized placeholders, ensuring deterministic and consistent replacement across time and modalities. Together, these components establish a secure interaction contract between the agent and the interface, decoupling task execution from direct access to sensitive data.

In summary, our contributions are as follows:
\begin{itemize}
    \item We propose a novel anonymization-based privacy protection framework for mobile GUI agents that enforces the principle of \emph{available-but-invisible} access to sensitive information.
    \item We design a type-preserving, deterministic anonymization mechanism and a secure interaction pipeline that allow agents to operate on sensitive data without observing its raw content.
    \item We address the consistency and grounding challenges introduced by anonymization and demonstrate that reliable agent execution can be maintained across modalities and time.
\end{itemize}

\section{Related Work}

\subsection{GUI Agents and Privacy Challenges}

GUI agents automate interactions on mobile and web platforms by processing visual (screenshots) or structural (XML/DOM) inputs through Large Language Models (LLMs) or Multimodal LLMs (MLLMs). Early works~\cite{mobileagent, autoglm, uitars2, appagent, appcopilot}, demonstrated significant potential in task automation but largely overlooked privacy risks. These agents often transmit full UI states—including unredacted screenshots or XML trees—to cloud-based models, which may expose Personally Identifiable Information (PII) such as contact details, financial records, or login credentials.

Recent surveys~\cite{guiagentssurvey1, guiagentssurvey2, osagentsurvey, guiagentsurvey} highlight that data privacy remains a key barrier to deploying GUI agents in real-world environments. The challenge is further amplified in multimodal settings, where visual encoders may capture sensitive information that text-only filters cannot detect~\cite{mindthethirdeye}. Moreover, modern agents rely on continuous screenshot streams and centralized routers or tool mediators~\cite{mcpagentbench, mcpworld, routerbench}, enabling correlation across time that facilitates user profiling and sensitive information reconstruction. Consequently, securing agent interactions without compromising utility has become an active research focus.

\subsection{PII Detection and Extraction in UIs}

Accurate detection and extraction of sensitive information is crucial for data security in GUI agents. Compared to standard NLP tasks, PII in UIs is sparse, context-dependent, and often multimodal. We categorize existing approaches as follows:

\begin{itemize}[leftmargin=0.2in]

    \item \textbf{Text-based Extraction.} Traditional methods rely on regular expressions or rule-based heuristics to identify patterns such as emails or phone numbers~\cite{guivulnerability}. These approaches lack semantic understanding and struggle with ambiguous data, e.g., distinguishing a bank balance from a generic numeric field. Modern Named Entity Recognition (NER) methods~\cite{gliner, glinermultitask, ai4p, microsoft_presidio, gliner_pii_large_v1}, employ bidirectional transformers to detect arbitrary entity types, offering better generalization for semantic PII detection.
    
    \item \textbf{Multimodal and Structural Extraction.} In GUIs, spatial layout and context are critical—for example, a number next to the label ``Balance'' is sensitive. Layout-aware models like LayoutLMv3~\cite{layoutlmv3} and UI-specific vision-language models such as ScreenAI~\cite{screenai}, VisionLLM~\cite{visionllmv2}, and Gemini~\cite{gemini25} integrate textual, visual, and layout information. These models enable precise localization and semantic inference of sensitive elements before they are processed by downstream reasoning modules. However, their strong representational capacity also enables more powerful privacy extraction, increasing the urgency of protection mechanisms.
    
\end{itemize}

\subsection{Privacy Protection Techniques in GUI Agents}

Existing methods for mitigating privacy risks in GUI agents generally fall into three categories:

\begin{itemize}[leftmargin=0.2in]

\item \textbf{Minimizing Exposure via Architecture.} CORE~\cite{core} addresses the ``all-or-nothing'' upload problem by partitioning the UI into semantic blocks based on the XML hierarchy. A local LLM filters content, sending only task-relevant blocks to a cloud LLM. This reduces exposure but does not anonymize the transmitted content, leaving any uploaded sensitive information fully visible.

\item \textbf{Adversarial and Visual Obfuscation.} A line of work explores adversarial perturbations that inject structured noise into images to disrupt vision-language models’ ability to recognize sensitive content~\cite{coattack, anyattack, foaattack, vip}. These methods optimize perturbations to break image-text alignment~\cite{coattack}, enhance transferability across models~\cite{foaattack}, or selectively perturb regions of interest~\cite{vip}. While effective in general VQA or captioning tasks, these approaches are not designed for the mobile agent setting, where the perturbation must simultaneously suppress privacy extraction while preserving fine-grained UI semantics for task execution.

DualTAP~\cite{dualtap} explicitly formulates this as a dual-objective problem, combining a privacy-interference loss with a task-preservation loss and introducing contrastive attention to localize privacy-sensitive regions. This allows targeted perturbation of sensitive areas while maintaining the agent’s operational accuracy. Unlike generic attacks, DualTAP is designed for continuous screenshot streams and agent workflows, making it more suitable for GUI-based environments.

\item \textbf{Anonymization and Intermediary Systems.} PrivWeb~\cite{privweb} injects a privacy layer that anonymizes PII in the DOM before reaching the agent. Using a local LLM, it categorizes data sensitivity and prompts users for authorization when necessary. While effective in text-based web environments, it cannot allow agents to directly use sensitive data (e.g., filling forms) without either revealing actual values or interrupting workflow, limiting its applicability to fully automated agents.
\end{itemize}

\section{Method}
\label{sec:method}

This section describes a tiered decoupling architecture designed to enable cloud-based multimodal GUI agents to perform high-level semantic reasoning while preventing exposure of raw sensitive data. The architecture places a secure local mediation layer between the device and any external reasoning service; this layer intercepts device state and user instructions, synthesizes a non-sensitive \emph{Virtual UI}, mediates all agent actions, and — when needed — performs tightly scoped local computations. Below we present the logical design, algorithmic building blocks, and the secure interaction protocols that realize this approach.

%  -------------------------------------------------

\subsection{System overview and design goals}
\label{sec:overview}

The core design principle of our system is \textbf{data anonymization} rather than data minimization. 
Instead of merely reducing the amount of transmitted data, our goal is to transform sensitive data into category-preserving anonymized representations such that the cloud agent can still perform semantic reasoning and task planning without accessing raw values.

\paragraph{Overall pipeline.}
The system follows a unified anonymization pipeline for both textual and visual modalities:

\begin{enumerate}
    \item Perform PII detection on textual inputs (user instructions and XML UI hierarchies) and on visual inputs (screenshots via OCR~\cite{easyocr}).
    \item Generate a category-preserving placeholder with a hash suffix for each detected PII entity.
    \item Replace the original textual spans or visual regions with the generated placeholders.
    \item Send only the anonymized Virtual UI to the cloud agent.
\end{enumerate}

Figure~\ref{fig:instruction_anon_example} illustrates an example of category-preserving anonymization applied to user instructions, where sensitive elements like names and phone numbers are replaced with placeholders.

\begin{figure}[h]
\centering
\fbox{
\begin{minipage}{0.95\linewidth}
\small
\textbf{Original instruction}
\begin{quote}\ttfamily
Add a contacts whose name is Xu, set the working phone number to be 12345678 and mobile phone number to be 87654321
\end{quote}

\textbf{Masked instruction}
\begin{quote}\ttfamily
Add a contacts whose name is \textbf{LAST\_NAME\#9zv3p}, set the working phone number to be \textbf{PHONE\_NUMBER\#cbnhu} and mobile phone number to be \textbf{PHONE\_NUMBER\#pilzc}
\end{quote}
\end{minipage}}
\caption{Example of category-preserving anonymization of user instructions.}
\label{fig:instruction_anon_example}
\end{figure}

\paragraph{Consistency challenges and their impact on agent execution.}
Although hash-based placeholders enable deterministic anonymization, they introduce consistency challenges across time and modalities that directly impair the agent’s ability to ground its actions in the interface. We identify two representative failure modes and illustrate how each leads to execution errors: 
1. \textbf{Inconsistent PII detection across modalities.}  
In some cases, the same entity is detected as PII in one modality but not in another, leading to a mismatch between real values and anonymized placeholders. Table~\ref{tab:inconsistent-detection} shows an example where the name ``Alice'' remains unmasked in the user instruction but is anonymized in the screenshot. As a result, the agent issues actions targeting ``Alice'', while the interface only exposes \texttt{FIRST\_NAME\#g7wef}, causing the agent to fail to locate the intended element.
2. \textbf{Inconsistent placeholder assignment for the same entity.}  
Even when an entity is consistently detected as PII, it may still be mapped to different anonymized identifiers at different times or in different modalities. 
Table~\ref{tab:inconsistent-placeholder} illustrates this issue with an example where the same entity ('Alice') is inconsistently assigned different placeholders across modalities. In this case, the agent searches for \texttt{FIRST\_NAME\#8dfa9} on the screen, but the interface presents \texttt{FIRST\_NAME\#g7wef}, again preventing successful grounding and execution.

\begin{table}[H]
\centering
\small
\caption{Example of inconsistent PII detection across modalities.}
\label{tab:inconsistent-detection}
\begin{tabular}{lll}
\toprule
\textbf{Source} & \textbf{Original} & \textbf{Masked} \\
\midrule
User instruction & Alice & Alice \\
Screenshot (OCR) & Alice & FIRST\_NAME\#g7wef \\
\bottomrule
\end{tabular}
\end{table}

\begin{table}[H]
\centering
\small
\caption{Example of inconsistent placeholder generation for the same entity.}
\label{tab:inconsistent-placeholder}
\begin{tabular}{lll}
\toprule
\textbf{Source} & \textbf{Original} & \textbf{Masked} \\
\midrule
User instruction & Alice & FIRST\_NAME\#8dfa9 \\
Screenshot 1 & Alice & FIRST\_NAME\#g7wef \\
Screenshot 2 & Alice & FIRST\_NAME\#8dfa9 \\
\bottomrule
\end{tabular}
\end{table}

Together, these examples demonstrate that cross-modality and temporal consistency is a necessary condition for reliable agent operation: the agent must observe exactly the same anonymized identifier for the same semantic entity throughout the task lifecycle. We summarize the consistency requirements as: 

\begin{itemize}
    \item \textbf{Problem A: Inconsistent PII detection}
    \begin{itemize}
        \item (A.1) An entity is detected as PII earlier but not later.
        \item (A.2) An entity is not detected as PII earlier but is detected later.
    \end{itemize}
    \item \textbf{Problem B: Inconsistent placeholder assignment} — the same entity is mapped to different placeholders.
\end{itemize}

\paragraph{Design objective.}
Our framework therefore enforces:
\begin{enumerate}
    \item Stable PII classification across time and modalities.
    \item Deterministic and globally consistent placeholder mapping within each task session.
\end{enumerate}

These objectives motivate the consistency enforcement and cross-modality synchronization mechanisms introduced in the following sections.

% ----------------------------------------------

\subsection{Logical layers}
Table~\ref{tab:layers} summarizes the four logical layers and their primary responsibilities.

\begin{table}[H]
  \centering
  \caption{Logical layers of the tiered decoupling architecture.}
  \label{tab:layers}
  \begin{tabular}{ll}
    \toprule
    \textbf{Layer} & \textbf{Primary responsibility} \\
    \midrule
    Layer 1: PII Detector & High-recall semantic identification of sensitive entities (NER + regex) \\
    Layer 2: UI Transformer & Deterministic pseudonymization, Virtual UI synthesis (XML + screenshot) \\
    Layer 3: Interaction Proxy & Interception, sanitization and reverse mapping of agent commands \\
    Layer 4: Privacy Gatekeeper & Policy-driven local computation for necessary semantic checks \\
    \bottomrule
  \end{tabular}
\end{table}

A session-scoped local mapping table is maintained within the trusted environment to record the correspondence between real entity values and their anonymized placeholders. This table is consulted by the PII detector, UI transformer, and interaction proxy to ensure deterministic placeholder assignment and consistent replacement across time and modalities. The mapping table never leaves the trusted boundary and raw values are never exposed to external services.

\subsection{Layer 1: PII detection and contextual whitelisting}
Layer~1 functions as the semantic gateway. It employs a hybrid detection strategy combining a label-guided, zero-shot NER model and deterministic pattern matchers.

\subsubsection{Label-guided zero-shot recognition}
The primary NER component uses a label-guided model from the GLiNER family~\cite{gliner_pii_large_v1} that accepts arbitrary label sets at inference time and supports custom entity detection with configurable sensitivity thresholds. This allows the system to explicitly define what constitutes sensitive information and to adjust the detection strictness according to different privacy or deployment requirements.
For a text input \(x\) (from prompts or XML text attributes), the detector produces a set of sensitive spans
\[
D = \{e_1, e_2, \dots, e_n\},
\]
where each span \(e_i\) is associated with a type \(T_i\) (e.g., \texttt{PHONE\_NUMBER}, \texttt{EMAIL}) and a confidence score \(S_i\). The zero-shot capability lets the system adapt the notion of sensitivity to different domains without re-training.

\subsubsection{Regex fallback and structural validation}
To capture low-context, high-entropy identifiers (e.g., credit-card numbers and standardized IDs), Layer~1 includes a deterministic regex layer. Regex detections complement the NER output and provide near-deterministic coverage for structured tokens that neural models may miss. 

In addition, we apply regex-based structural validation to XML inputs with an explicit whitelist of XML-specific keywords and structural tokens. This exemption mechanism prevents fundamental XML representations (such as tag names, attribute keys, and schema-related identifiers) from being mistakenly classified as PII, thereby reducing false positives introduced by pattern-based matching on semi-structured data.

\subsubsection{Instruction-driven contextual whitelisting}
We construct a session-scoped whitelist \(W\) derived from the user's instruction \(I\) to mitigate inconsistency in PII detection across different modalities and contexts (Problem~A.2). The core principle is that if a token is explicitly mentioned in the user instruction and is classified as non-PII in that context, it will be treated as non-sensitive throughout the session and exempted from anonymization in subsequent XML or screenshot-derived text.

This design is motivated by two observations. First, neural PII detectors are inherently unstable and context-sensitive: the same string may be classified differently depending on surrounding text, modality (natural language vs.\ UI structure vs.\ OCR output), or minor perturbations. Second, user instructions are typically more natural, fluent, and semantically coherent than UI representations or OCR-extracted text, making them a more reliable context for determining whether a string constitutes PII. Moreover, if a truly sensitive string is already exposed in the user instruction and classified as non-PII, anonymizing it later in XML or screenshots is ineffective, as the privacy leakage has already occurred.

The whitelisting procedure is:

\begin{algorithm}[H]
\caption{Instruction-Driven Contextual Whitelist Construction}
\begin{algorithmic}[1]
\State \textbf{Input:} user instruction \(I\)
\State \(E_{\text{prompt}} \leftarrow\) PII\_Detector(\(I\)) \Comment{Detected sensitive entities in the prompt}
\State \(K \leftarrow\) extract\_functional\_tokens(\(I, E_{\text{prompt}}\)) \Comment{Tokens classified as non-PII,}
\State \(W \leftarrow W \cup K\) \Comment{Add non-sensitive instruction tokens to session whitelist}
\State \Return \(W\)
\end{algorithmic}
\end{algorithm}

During subsequent UI scans, any candidate token \(t \in W\) will bypass anonymization and be consistently treated as non-PII.

\subsection{Layer 2: UI transformer and Virtual UI synthesis}

\begin{figure}[htbp]
    \centering
    \includegraphics[width=0.6\textwidth]{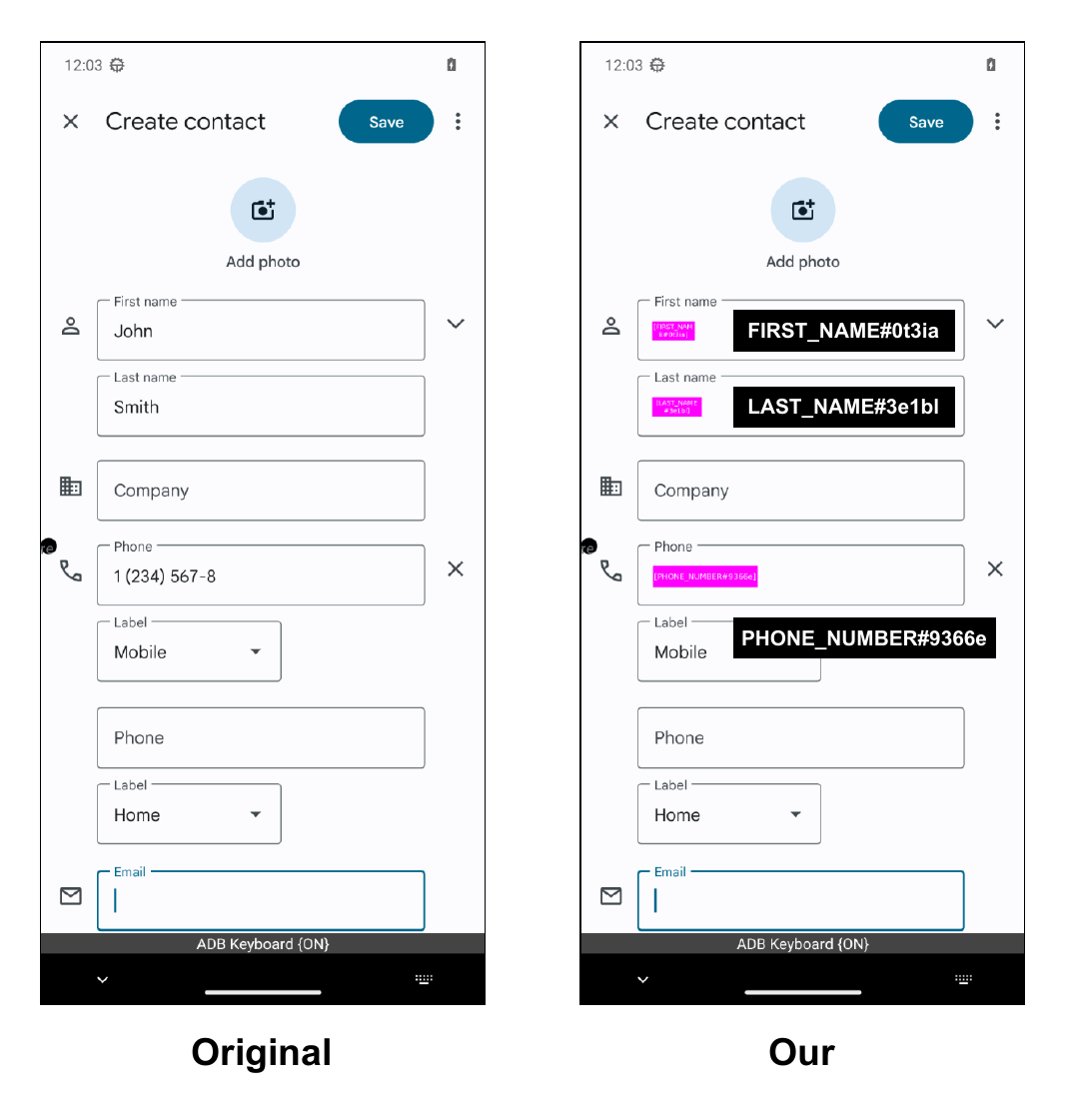}
    \caption{Comparison of screenshots before and after anonymization. 
The left image shows the original screen before anonymization, while the right image illustrates the anonymized version. 
The text in black regions highlights enlarged excerpts of the magenta regions to illustrate the corresponding content.}
    \label{fig:virtualUI}
\end{figure}

Layer~2 performs deterministic pseudonymization and produces the \emph{Virtual UI} consumed by the cloud agent. Its goals are to preserve semantics (type, spatial location, interactivity) while removing raw sensitive values, as demonstrated in Figure~\ref{fig:virtualUI}, which compares original and anonymized screenshots.

\subsubsection{Deterministic placeholder generation}
For each detected entity value \(v\) of type \(T\), the system generates a stable placeholder \(P\) using a cryptographic-hash-based procedure to maintain session consistency and collision resistance:
\[
P = T\#\operatorname{Truncate}\big(\operatorname{Base36}(\operatorname{SHA256}(v \,\|\, T)),\ 5\big),
\]
where ``\(\|\)'' denotes string concatenation, \(\operatorname{Base36}\) encodes the hash digest into Base36, and \(\operatorname{Truncate}(\cdot,5)\) retains the leading 5 characters. Concatenating the type \(T\) ensures identical raw strings used in different semantic roles map to distinct placeholders.

\subsubsection{Cross-modality consistency}
This mechanism is designed to mitigate both \textbf{Problem A.1} (missed or inconsistent PII detection) and \textbf{Problem B} (inconsistent placeholder assignment across contexts and modalities). 

At the core of this mechanism is a local \textbf{mapping table}, which maintains the mapping from each real entity value to its anonymized placeholder. 
This ensures that all appearances of the same entity—across different modalities or locations—are consistently represented by the same placeholder.

The transformer consults this table and enforces a prioritized cross-modality alignment strategy to ensure consistent placeholder usage:

\begin{itemize}[leftmargin=0.2in]
  \item \textbf{Instruction-derived mappings have top priority.}  
  After performing PII detection on the user instruction, all detected entities are immediately assigned anonymized placeholders and inserted into the local mapping table.

  \item \textbf{XML text attributes} (e.g., \texttt{text}, \texttt{hint}, \texttt{content-desc}) are anonymized by first querying the mapping table.  
  If an extracted entity value already exists in the table, the corresponding placeholder is reused; otherwise, a new placeholder is generated and added to the table.

  \item \textbf{OCR-extracted strings from screenshots} are normalized and fuzzy-matched against entity values already registered in the mapping table. 
  If a match is found, the corresponding placeholder is reused; otherwise, the string is treated as a new entity and handled by the same lookup-or-create mechanism.
\end{itemize}

This lookup-before-generation policy ensures that semantically identical entities appearing at different locations or in different modalities are always mapped to the same placeholder, thereby alleviating \textbf{Problem B}. 
Moreover, even if an entity is missed by the PII detector in one modality, it can still be replaced through matching with previously recorded values in the mapping table, which partially mitigates \textbf{Problem A.1}.

\subsubsection{OCR pipeline and fuzzy alignment}
Screenshot processing uses an OCR~\cite{easyocr} pipeline (text detection, recognition, decoding). Extracted OCR strings are chunked and fed to the detector. Because OCR output can be noisy, the system uses normalized Levenshtein distance to compute similarity between an OCR string \(s_2\) and a registered entity value \(s_1\) stored in the mapping table:
\[
R(s_1, s_2) = 1 - \frac{\operatorname{Levenshtein}(s_1, s_2)}{\max(|s_1|, |s_2|)}.
\]
If \(R > \tau\) (we use \(\tau=0.85\) in our implementation), \(s_2\) is mapped to the placeholder associated with \(s_1\). This fuzzy alignment reduces grounding failures caused by OCR errors.

The accuracy of the OCR model can influence the overall task success rate, as it directly affects the reliability of PII detection in screenshots. Since OCR is invoked at every step of the task execution, we recommend opting for a lightweight and faster OCR model to ensure efficiency and minimize latency. Nevertheless, the choice can be tailored to the specific hardware capabilities available, allowing for the selection of more advanced models when computational resources permit.

\subsubsection{Visual masking and placeholder rendering}
Visual masking is implemented by drawing opaque overlays over sensitive bounding boxes and rendering the corresponding placeholder text within the masked region. The overlay color and text rendering are selected to preserve spatial cues for the VLM while preventing recovery of the underlying pixels. Text is scaled to fit the bounding box so the VLM can infer category and approximate value length without seeing raw data.

\subsection{Layer 3: Secure interaction proxy}
The Secure Interaction Proxy mediates all agent-issued actions and is responsible for reverse mapping, coordinate resolution, and enforcement of execution policies.

Our command interface is built on the design of AndroidLab~\cite{androidlab}, which maps interactable UI elements from the XML hierarchy to numeric indices on screenshots. 
We extend and modify this design to support additional gesture types and flexible text handling. 
Every agent action (e.g., \texttt{tap(num)} or \texttt{type(str)}) is intercepted by the local proxy before execution. 
The proxy resolves and validates commands according to their parameter types and semantics:

\begin{itemize}[leftmargin=0.2in]
  \item For spatial actions such as \texttt{tap}, \texttt{long\_press}, and \texttt{swipe}, the agent issues a numeric index (\texttt{num}) referring to an interactable UI element annotated on the screenshot. The proxy verifies that the index is within range, maps it to the corresponding bounding box extracted from the XML hierarchy, and converts it into concrete device coordinates (centroid) before execution.
  \item For textual actions such as \texttt{type}, the parameter is always a string. This string may consist solely of anonymized placeholders, solely of natural language text, or a mixture of both. The proxy identifies and resolves all placeholders within the string to their corresponding raw values, while leaving any natural language content unchanged.
  \item The proxy rejects commands if the element list is empty or the referenced index is out of range, preventing execution on stale or inconsistent UI states.
\end{itemize}

\begin{figure}[htbp]
    \centering
    \includegraphics[width=0.8\textwidth]{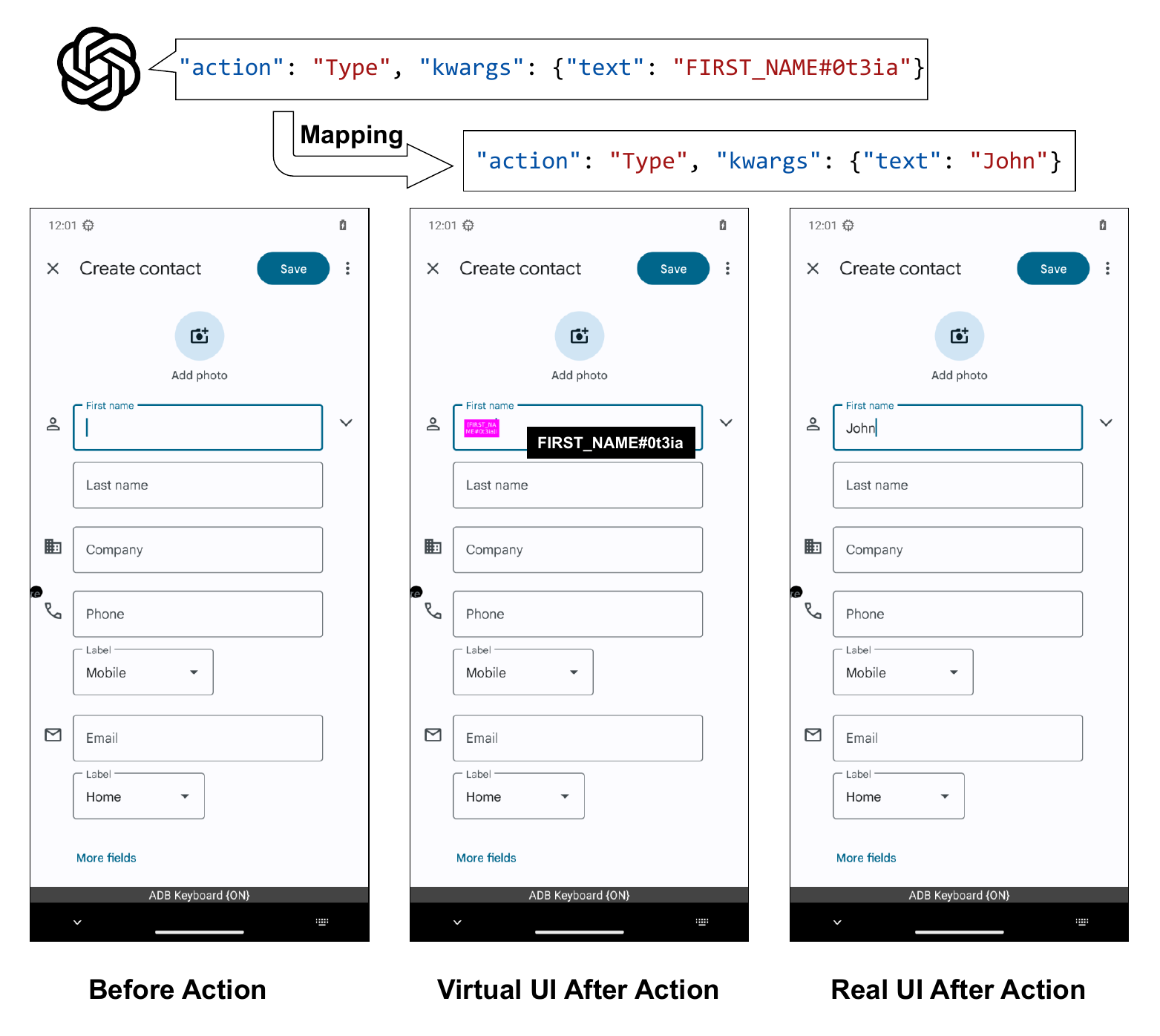}
    \caption{Example of \textbf{Type} proxy resolution. The text in black regions highlights enlarged excerpts of the magenta regions to illustrate the corresponding content.}
    \label{fig:type_resolution}
\end{figure}

Importantly, only \texttt{type} parameters contain textual content that may include placeholders. 
All spatial and gesture-based commands operate purely on numeric indices and coordinates and therefore never include anonymization tags. Table~\ref{tab:proxy} summarizes representative command types and their resolution logic, while Figure~\ref{fig:type_resolution} provides a visual example of how the 'Type' command's argument is resolved through placeholder mapping.

\begin{table}[htbp]
  \centering
  \small
  \caption{Representative proxy resolution logic for common commands.}
  \label{tab:proxy}
  \begin{tabular}{p{0.13\linewidth}p{0.25\linewidth}p{0.53\linewidth}}
    \toprule
    Action & Agent form & Proxy resolution \\
    \midrule
    Tap & \texttt{tap(num)} & validate index; map bbox centroid $(x,y)$; execute tap \\
    Long press & \texttt{long\_press(num)} & validate index; map bbox centroid; execute long press \\
    Swipe & \texttt{swipe(num, dir, dist)} & validate index; map centroid; execute swipe$(dir,dist)$ \\
    Type  & \texttt{type(str)} & identify and replace any placeholders within the string; insert resolved text \\
    \bottomrule
  \end{tabular}
\end{table}

\subsection{Layer 4: Privacy Gatekeeper and secure local computation}
Some tasks require operations on raw sensitive values (e.g., comparing two dates, validating a payment amount). Layer~4 provides a narrow, auditable local computation interface managed by a small local language model (SLM) that acts as a \emph{Privacy Gatekeeper}. In our implementation, this SLM is powered by the Qwen3-8B causal language model~\cite{qwen3technicalreport}, which supports both reasoning and instruction-following capabilities and can efficiently handle the necessary local computation steps.

When the cloud agent needs a local computation, it issues a structured request:
\[
\texttt{cloud\_agent\_compute\_with\_tokens(tokens, instruction, reason)}.
\]
The gatekeeper evaluates the request against three policy criteria:
\begin{itemize}[leftmargin=0.2in]
  \item \textbf{Relevance:} Is the computation necessary for the user’s stated goal?
  \item \textbf{Necessity:} Can the computation be satisfied using the Virtual UI alone, or does it require raw values?
  \item \textbf{Minimization:} Can the result be expressed as a bounded, non-revealing value (e.g., boolean, categorical label) rather than returning raw data?
\end{itemize}
Only when these checks pass does the SLM perform the computation on local raw values and return a minimal result (for example, \texttt{True}, \texttt{"greater\_than"}, or a categorical label). Raw values never leave the trusted boundary.

\subsection{Privacy governance and utility--privacy tradeoffs}
The architecture achieves data minimization through layered defenses and a local gatekeeper. Table~\ref{tab:tradeoff} summarizes representative features, their privacy benefits, and impacts on agent utility.

\begin{table}[h]
  \centering
  \caption{Selected features and tradeoffs (summary).}
  \label{tab:tradeoff}
  \begin{tabular}{lll}
    \toprule
    Feature & Privacy benefit & Utility impact \\
    \midrule
    Deterministic hashing & Prevents trivial linkage & Maintains entity identity across steps \\
    Contextual whitelisting & Avoids over-redaction & Preserves UI affordances \\
    Local secure computation & Raw data never leaves device & Enables comparisons/validations \\
    Fuzzy matching & Robustness to OCR noise & Reduces grounding errors \\
    \bottomrule
  \end{tabular}
\end{table}

\section{Experiment}

\subsection{Experimental Setup}

To evaluate the effectiveness of our Privacy Protection Layer in balancing privacy preservation and task utility for mobile agents, we conducted comprehensive experiments. All evaluations were performed within the AndroidLab framework~\cite{androidlab}, utilizing its default Android Virtual Device (AVD) setup on a Pixel 7 Pro emulator. 

\paragraph{Benchmarks and Tasks}
We utilized two distinct benchmarks: the \textbf{AndroidLab}~\cite{androidlab} benchmark and the \textbf{PrivScreen} benchmark provided by DualTAP~\cite{dualtap}. These serve complementary purposes, assessing general-purpose mobile tasks and privacy-specific scenarios respectively. The AndroidLab benchmark consists of 138 tasks (Operation and Query types) gathered from nine apps. It supports two agent modes: \textbf{XML Mode}, which uses a redesigned XML compression algorithm for text-only inputs, and \textbf{SoM (Set-of-Mark) Mode}, designed for multimodal models (LMMs) where clickable elements are assigned serial numbers. The PrivScreen benchmark is a dual-task QA-style dataset comprising over 500 real application screenshots augmented with 1000+ synthetic PII. It includes: (i) a privacy-focused QA to assess leakage, and (ii) a utility-focused QA to evaluate general functionality. We applied our anonymization to these images and ran the standard evaluation scripts.

\paragraph{Models and Baselines}
For \textbf{AndroidLab}, we employed two cloud-based models: \textbf{GPT-4o} and \textbf{Gemini-3-flash-preview}. Besides the unprotected input (\emph{Original}) and our anonymization pipeline (\emph{Ours}), we include a strong privacy baseline, \emph{Full cover}, which replaces all detected sensitive spans with a fixed token (\texttt{[privacy information]}) instead of category-preserving placeholders. In \textbf{SoM} mode, we further include the screenshot perturbation method from DualTAP~\cite{dualtap} as an additional baseline (\emph{DualTAP}); this baseline is not applicable to \textbf{XML} mode because XML mode does not consume screenshot inputs. For \textbf{PrivScreen}, we evaluated three multimodal models: \textbf{Qwen2.5-VL-7B}~\cite{qwen25vl}, \textbf{GPT-5}, and \textbf{UI-TARS-7B}~\cite{uitars}. Our method (\emph{Ours}) is compared against several baselines from the DualTAP study: \emph{Original} (undefended), \emph{AnyAttack}~\cite{anyattack}, \emph{FOA-Attack}~\cite{foaattack}, \emph{VIP}~\cite{vip}, and \emph{DualTAP}~\cite{dualtap}. Note that the baseline results on PrivScreen are cited directly from the DualTAP paper.

\paragraph{Metrics}
On \textbf{AndroidLab}, we employ: \textbf{Success Rate (SR)}, \textbf{Sub-Goal Success Rate (Sub-SR)}, \textbf{Reversed Redundancy Ratio (RRR)} for efficiency, and \textbf{Reasonable Operation Ratio (ROR)} to measure screen-change effectiveness. \textbf{The evaluation protocol for SR depends on task type}. For operation tasks, we arrange all execution screenshots in chronological order and provide them with the task description to a vision model, which judges whether the task is successfully completed; the vision evaluator is \textbf{Qwen3-VL-235B-A22B-Thinking}. For \texttt{query\_detect} tasks, we provide the task description, reference answer, and the GUI agent's answer to a text model, which judges whether the response is correct; the text evaluator is \textbf{DeepSeek-V3.2-Instruct}. \textbf{Sub-SR, RRR, and ROR} are still computed with AndroidLab's original automatic evaluation pipeline. On \textbf{PrivScreen}, utility is measured by \textbf{Accuracy (Acc)}. Privacy leakage is assessed at the character level via \textbf{Leakage Rate (LR)}, \textbf{Match Score (MS)}, \textbf{BLEU}, and \textbf{ROUGE-L}. Semantic-level leakage is measured via \textbf{BERTScore} and \textbf{Cosine Similarity (CS)}.

\paragraph{Implementation Details} 
The Privacy Protection Layer is integrated into the AndroidLab pipeline at four points: prompt anonymization, XML anonymization, screenshot masking (using OCR and magenta-background overlays), and de-anonymization (token-to-real conversion) during local execution. For PII detection, we used \textbf{EasyOCR}~\cite{easyocr} for text extraction and \textbf{GLiNER} (\texttt{gliner-pii-large-v1.0}~\cite{gliner_pii_large_v1}) for Named Entity Recognition (threshold 0.5) with regex fallbacks. Experiments were conducted on an \textbf{NVIDIA GeForce RTX 4090 GPU}, with the NER model occupying approximately 2800 MB of VRAM.

\subsection{Main Results}

\begin{table*}[t]
\centering
\small
\caption{AndroidLab benchmark}
\label{tab:andlab}
\begin{tabular}{lllcccc}
\toprule
\textbf{Agent Mode} & \textbf{Model} & \textbf{Protection} & \textbf{SR} $\uparrow$ & \textbf{Sub-SR} $\uparrow$ & \textbf{RRR} $\uparrow$ & \textbf{ROR} $\uparrow$ \\ 
\midrule

\multirow{8}{*}{SoM} 

 & \multirow{4}{*}{GPT-4o} 
  & Original & \textbf{44.20} & \textbf{81.08} & \textbf{90.00} & \textbf{89.34} \\
 & & Full cover & 34.78 & 76.18 & 86.16 & 87.48 \\
 & & DualTAP & 33.33 & 71.65 & 72.16 & 85.77 \\
 & & Ours     & 40.58 & 79.43 & 80.47 & 86.68 \\

\cmidrule(lr){2-7}

 & \multirow{4}{*}{Gemini-3-flash-preview} 
  & Original & \textbf{70.29} & \textbf{88.45} & \textbf{87.50} & \textbf{94.43} \\
 & & Full cover & 55.80 & 82.84 & 85.75 & 92.32 \\
 & & DualTAP & 63.77 & 85.59 & 82.62 & 92.72 \\
 & & Ours     & 63.77 & 85.43 & 87.04 & 93.73 \\

\midrule

\multirow{6}{*}{XML} 

 & \multirow{3}{*}{GPT-4o} 
  & Original & 35.51 & 74.30 & 92.40 & 90.12 \\
 & & Full cover & 39.86 & 76.07 & \textbf{93.62} & \textbf{90.84} \\
 & & Ours     & \textbf{41.30} & \textbf{77.36} & 92.15 & 89.55 \\

\cmidrule(lr){2-7}

 & \multirow{3}{*}{Gemini-3-flash-preview} 
  & Original & \textbf{57.97} & \textbf{85.28} & 91.86 & \textbf{93.73} \\
 & & Full cover & 50.00 & 83.49 & \textbf{94.07} & 92.85  \\
 & & Ours     & 54.34 & 84.32 & 90.91 & 92.67 \\

\bottomrule
\end{tabular}
\end{table*}

\paragraph{Results on AndroidLab.}
The main results on the AndroidLab benchmark are reported in Table~\ref{tab:andlab}. Overall, our method introduces only limited utility degradation after privacy protection, and in most settings clearly outperforms \emph{Full cover} and \emph{DualTAP}, showing that category-preserving anonymization provides a favorable privacy--utility trade-off.

Under SoM mode, the unprotected setting achieves the highest SR and Sub-SR for both backbones, but our method remains notably closer to \emph{Original} than the two baselines. For GPT-4o, \emph{Original} reaches 44.20\% SR, while \emph{Full cover} and \emph{DualTAP} drop to 34.78\% and 33.33\%; our method still achieves 40.58\%, with Sub-SR also higher than both baselines. For Gemini-3-flash-preview, our SR is 63.77\%, matching \emph{DualTAP} and clearly above \emph{Full cover} (55.80\%), while maintaining a high ROR of 93.73\%. This suggests that our method preserves more stable interaction behavior under multimodal inputs.

Under XML mode, the methods exhibit a finer-grained trade-off. For GPT-4o, our method achieves the best SR and Sub-SR (41.30\%/77.36\%), outperforming both \emph{Original} and \emph{Full cover}, while \emph{Full cover} remains best on RRR and ROR. For Gemini-3-flash-preview, \emph{Original} is still strongest on SR, Sub-SR, and ROR, and \emph{Full cover} is best on RRR; our method lies between them overall, retaining strong task completion while avoiding the information loss caused by blanket replacement.

\begin{table}[htbp]
\centering
\small
\caption{PrivScreen benchmark}
\label{tab:dualtap}
\begin{tabular}{ll ccccccc}
\toprule
\textbf{Model} & \textbf{Protection Method} & \textbf{Acc} $\uparrow$ & \textbf{LR} $\downarrow$ & \textbf{MS} $\downarrow$ & \textbf{BertScore} $\downarrow$ & \textbf{CS} $\downarrow$ & \textbf{BLEU} $\downarrow$ & \textbf{ROUGE-L} $\downarrow$ \\

\midrule

\multirow{5}{*}{Qwen2.5-VL-7B} 
& Original   & 89.00 & 97.14 & 96.99 & 0.8675 & 0.9011 & 0.7830 & 0.8991 \\
& AnyAttack  & 70.00 & 95.24 & 91.86 & 0.6721 & 0.8705 & 0.5164 & 0.7317 \\
& FOA-Attack & 73.00 & 83.81 & 82.84 & 0.6307 & 0.8150 & 0.5343 & 0.7309 \\
& VIP        & 81.00 & 94.76 & 92.67 & 0.7556 & 0.8960 & 0.5932 & 0.7856 \\
& DualTAP    & 88.00 & 32.86 & 38.82 & 0.1998 & 0.4202 & \textbf{0.0908} & \textbf{0.1660} \\
& \textbf{Ours} & \textbf{89.00} & \textbf{19.52} & \textbf{30.69} & \textbf{0.1411} & \textbf{0.3966} & 0.1496 & 0.1832 \\

\midrule

\multirow{6}{*}{GPT-5} 
& Original   & 93.00 & 97.14 & 97.15 & 0.9260 & 0.9627 & 0.8584 & 0.9246 \\
& AnyAttack  & 90.00 & 91.43 & 87.49 & 0.6757 & 0.8490 & 0.4978 & 0.7446 \\
& FOA-Attack & 80.00 & 80.95 & 79.72 & 0.6293 & 0.7532 & 0.4967 & 0.7004 \\
& VIP        & 86.00 & 80.00 & 79.33 & 0.6662 & 0.7715 & 0.5284 & 0.6846 \\
& DualTAP    & \textbf{91.00} & 23.33 & \textbf{24.43} & 0.1589 & \textbf{0.2379} & \textbf{0.0941} & \textbf{0.1628} \\
& \textbf{Ours} & 90.00 & \textbf{18.84} & 26.74 & \textbf{0.1495} & 0.3224 & 0.1587 & 0.1892 \\

\midrule

\multirow{5}{*}{UI-TARS-7B} 
& Original   & 80.00 & 93.81 & 93.10 & 0.7846 & 0.9079 & 0.6721 & 0.8363 \\
& AnyAttack  & \textbf{76.00} & 81.43 & 80.11 & 0.5533 & 0.7615 & 0.3748 & 0.6057 \\
& FOA-Attack & 64.00 & 76.67 & 75.60 & 0.5349 & 0.7187 & 0.4229 & 0.6178 \\
& VIP        & 66.00 & 89.05 & 85.83 & 0.6739 & 0.8288 & 0.5232 & 0.7223 \\
& DualTAP    & 67.00 & 34.76 & 38.35 & 0.2278 & 0.4535 & 0.1139 & 0.2034 \\
& \textbf{Ours} & 61.00 & \textbf{18.10} & \textbf{29.15} & \textbf{0.1251} & \textbf{0.3695} & \textbf{0.1029} & \textbf{0.1552} \\

\bottomrule
\end{tabular}
\end{table}

% \midrule

% \multirow{5}{*}{Gemini-2.0 flash} 
% & Original   & 87.00 & 96.67 & 96.72 & 0.8651 & 0.9574 & 0.8030 & 0.9302 \\
% & AnyAttack  & 82.00 & 97.14 & 96.66 & 0.7993 & 0.9400 & 0.6960 & 0.9001 \\
% & FOA-Attack & 79.00 & 94.76 & 94.13 & 0.7401 & 0.9045 & 0.6362 & 0.8342 \\
% & VIP        & 84.00 & 94.79 & 92.76 & 0.7311 & 0.8830 & 0.5746 & 0.7828 \\
% & DualTAP    & 87.00 & 57.62 & 61.41 & 0.3805 & 0.5193 & 0.1686 & 0.2592 \\

% \midrule

% \multirow{5}{*}{Holo1.5-7B} 
% & Original   & 70.00 & 94.76 & 94.07 & 0.8351 & 0.9036 & 0.7182 & 0.8232 \\
% & AnyAttack  & 61.00 & 81.43 & 79.07 & 0.5711 & 0.7447 & 0.3886 & 0.5801 \\
% & FOA-Attack & 58.00 & 78.10 & 78.29 & 0.5590 & 0.7571 & 0.3964 & 0.5852 \\
% & VIP        & 57.00 & 89.05 & 86.19 & 0.6559 & 0.8263 & 0.4633 & 0.6583 \\
% & DualTAP    & 69.00 & 17.14 & 30.75 & 0.1327 & 0.3620 & 0.0302 & 0.0949 \\

% \midrule
% \multirow{5}{*}{InternVL3-5-8B} 
% & Original   & 83.00 & 96.19 & 92.96 & 0.7172 & 0.8866 & 0.5586 & 0.7770 \\
% & AnyAttack  & 78.00 & 90.48 & 86.38 & 0.5689 & 0.7671 & 0.3512 & 0.5702 \\
% & FOA-Attack & 70.00 & 78.57 & 78.54 & 0.5706 & 0.7198 & 0.4229 & 0.6012 \\
% & VIP        & 80.00 & 85.31 & 83.52 & 0.5801 & 0.8089 & 0.4022 & 0.6273 \\
% & DualTAP    & 83.00 & 24.29 & 38.16 & 0.1905 & 0.3996 & 0.0841 & 0.1275 \\

\paragraph{Results on PrivScreen.}
Table~\ref{tab:dualtap} presents the results on the PrivScreen benchmark. Overall, our method provides strong privacy protection across all tested models while keeping utility competitive. For Qwen2.5-VL-7B, our method maintains the same task accuracy (89.00\%) while dramatically reducing the LR from 97.14\% to 19.52\% and MS from 96.99\% to 30.69\%. For GPT-5, we preserve high utility (90.00\% vs. 93.00\%) while reducing BertScore from 0.9260 to 0.1495. Compared to existing defenses such as AnyAttack or DualTAP, our method consistently improves key leakage-related metrics while maintaining competitive accuracy.

\subsection{Overhead and PII Analysis}

We evaluate both the computational overhead and the effectiveness of our anonymization framework in handling sensitive information.

\begin{table}[htbp]
\centering
\caption{Average inference time of the privacy protection layer. We report the average per-image latency for each component. OCR (EasyOCR) is executed on CPU, while NER (GLiNER) is executed on an NVIDIA GeForce RTX 4090 GPU.}

\label{tab:overhead}
\begin{tabular}{lc}
\toprule
\textbf{Processing Step} & \textbf{Average Time per Image (seconds)} \\
\midrule
OCR (EasyOCR) & 0.838 \\
NER (GLiNER)  & 0.663 \\
\midrule
\textbf{Total Time} & \textbf{1.770} \\
\bottomrule
\end{tabular}
\end{table}

\begin{table}[htbp]
\centering
\small
\caption{Percentage of content detected as PII by our anonymization framework on the AndroidLab benchmark. Each value represents the ratio between the length of content identified as PII and the total input length for each task type.}
\label{tab:pii_ratio}
\begin{tabular}{lc}
\toprule
\textbf{Task Type} & \textbf{Percentage of content identified as PII} \\
\midrule
Task description & 3.49\% \\
Image & 4.38\% \\
XML & 0.42\% \\
\bottomrule
\end{tabular}
\end{table}

\paragraph{Computational Overhead.}
Table~\ref{tab:overhead} reports the average inference time per image for each component of the privacy protection layer. The OCR module (EasyOCR) takes 0.838 seconds on CPU, while the NER module (GLiNER) takes 0.663 seconds on an NVIDIA GeForce RTX 4090 GPU. The total additional latency introduced by the privacy protection layer is approximately 1.77 seconds per image. This overhead is moderate and acceptable in the context of mobile GUI automation, where tasks involve relatively long interaction sequences and are typically not latency-critical.

\paragraph{Proportion of Content Identified as PII.}
We further analyze the fraction of input content flagged as PII by our anonymization framework. Table~\ref{tab:pii_ratio} reports the percentage of content length detected as PII for different task types in the AndroidLab benchmark. As shown, only a small portion of the total input is considered sensitive: 3.49\% for task descriptions, 4.38\% for images, and 0.42\% for XML inputs. This indicates that the framework selectively targets sensitive information without significantly modifying the majority of the input content.

\section{Conclusion}
In this work, we presented a novel anonymization-based privacy protection framework for mobile GUI agents that enforces the principle of making sensitive information \emph{available but invisible}. Our framework enables agents to leverage sensitive data for task execution without directly observing raw values, through deterministic, type-preserving placeholders, a secure interaction proxy, and a local Privacy Gatekeeper. It consists of four coordinated components: (i) a PII Detector that achieves high-recall semantic identification of sensitive entities via a hybrid strategy combining label-guided zero-shot NER with regex-based fallback; (ii) a UI Transformer that performs deterministic pseudonymization and Virtual UI synthesis while preserving layout and interaction affordances; (iii) a Secure Interaction Proxy that intercepts, sanitizes, and reverse-maps agent commands; and (iv) a local Privacy Gatekeeper that executes policy-driven computations over raw values and returns only non-sensitive results.

A key conceptual contribution of this work is the distinction between data \emph{usage} and data \emph{understanding}. We characterize \emph{usage} as scenarios in which replacing a sensitive value alters only the input content without affecting the agent's control logic, whereas \emph{understanding} refers to cases where the specific semantic content is essential for decision-making. By combining anonymization with local semantic augmentation, our framework effectively supports both categories while maintaining strong privacy guarantees. Importantly, the layered design ensures cross-modality and temporal consistency in PII detection and placeholder assignment, which is critical for reliable multi-step GUI agent execution in dynamic mobile environments.

Extensive experiments on the AndroidLab and PrivScreen benchmarks demonstrate that our approach consistently reduces privacy leakage across multiple models, while incurring only modest utility degradation. These results indicate a strong and favorable privacy--utility trade-off compared to existing methods, validating the practicality of our framework for real-world mobile GUI agent scenarios.

\section{Discussion}
Beyond its technical implementation, this work highlights fundamental limitations in current GUI agent architectures. The prevailing reliance on raw visual screen content reflects a pragmatic but structurally constrained design choice, stemming from the absence of native, LLM-friendly interaction protocols (e.g., Tool Calling or Model Context Protocol) within mobile operating systems and third-party applications. This paradigm inherently exposes entire screen contents to agents, leading to systematic over-collection of sensitive information. Our findings suggest that this limitation is not merely an implementation artifact, but a structural property of today’s mobile OS ecosystems, and that achieving strong privacy guarantees in the long term may require a paradigm shift toward semantically structured and privacy-aware agent interfaces.

Within the constraints of existing platforms, our framework addresses several challenges, including cross-modality alignment and temporal consistency in sensitive information handling. Mechanisms such as session-scoped local mapping tables and instruction-driven contextual whitelisting mitigate grounding failures and ensure stable agent behavior across multi-step interactions. Nevertheless, tasks requiring complex semantic understanding over sensitive content remain challenging, pointing to important directions for future research. Promising avenues include the integration of on-device multimodal models, more expressive local semantic reasoning modules, and privacy-preserving computation interfaces that allow cloud-based agents to operate over protected or encrypted data, further advancing secure and practical agentic workflows.

%Bibliography
\bibliographystyle{unsrt}  
\bibliography{references}

\end{document}